\documentclass[iop]{emulateapj}
\usepackage{epsfig}

\usepackage[usenames,dvips]{color}

\shorttitle{NuSTAR Observation of PSR\,J1723--2837}
\shortauthors{Kong et al.}

\newcommand{\chandra}{{\it Chandra}}

\newcommand{\xmm}{{\it XMM-Newton}}
\newcommand{\swift}{{\it Swift}}
\newcommand{\fermi}{{\it Fermi}}
\newcommand{\nustar}{{\it NuSTAR}}
\newcommand{\msp}{PSR\,J1723--2837}

\begin{document}
\slugcomment{Accepted for publication in ApJ}
\title{A NuSTAR Observation of the Gamma-ray Emitting Millisecond Pulsar PSR\,J1723--2837}

\author{A.~K.~H. Kong\altaffilmark{1,2}, C.~Y. Hui\altaffilmark{3}, J. Takata\altaffilmark{4}, K.~L. Li\altaffilmark{5}, P.~H.~T. Tam\altaffilmark{6}
}

\affil{$^1$ Institute of Astronomy and Department of Physics, National Tsing Hua University, Hsinchu 30013, Taiwan; akong@phys.nthu.edu.tw}
\affil{$^2$ Astrophysics, Department of Physics, University of Oxford, Keble Road, Oxford OX1 3RH, UK}
\affil{$^3$ Department of Astronomy and Space Science, Chungnam National University, Daejeon, Republic of Korea; cyhui@cnu.ac.kr} 
\affil{$^4$ Institute of Particle Physics and Astronomy, Huazhong University of Science and Technology, China }
\affil{$^5$ Department of Physics and Astronomy, Michigan State University, East Lansing, MI 48824, USA} 
\affil{$^6$ Institute of Astronomy and Space Science, Sun Yat-Sen University, Guangzhou 510275, China} 




\begin{abstract}
We report on the first \nustar\ observation of the gamma-ray emitting millisecond pulsar binary \msp. X-ray radiation up to 79 keV is clearly detected and the simultaneous \nustar\ and \swift\ spectrum is well described by an absorbed power-law with a photon index of $\sim 1.3$. We also find X-ray modulations in the 3--10 keV, 10--20 keV, 20--79 keV, and 3--79 keV bands at the 14.8-hr binary orbital period. All these are entirely consistent with previous X-ray observations below 10 keV. This new hard X-ray observation of \msp\ provides strong evidence that the X-rays are from the intrabinary shock via an interaction between the pulsar wind and the outflow from the companion star. We discuss how the \nustar\ observation constrains the physical parameters of the intrabinary shock model.
\end{abstract}


\section{Introduction}
Thanks to the combined effort of continuous $\gamma-$ray all-sky observation
and radio pulsation searches, the population of millisecond pulsars (MSPs) 
are expanded significantly in the recent years (see the review by Hui 2014).  
Among these newly discovered MSPs, one interesting new class 
which is usually referred as ``redbacks" has emerged. They are characterized 
by their non-degenerate companions with the mass ranges from $\sim0.2M_{\odot}$ 
to $\sim0.7M_{\odot}$ and with the orbital period $P_{b}\lesssim20$~hrs (see Roberts 2013 for a review). 

The first discovered redback MSP PSR\,J1023+0038 has provided us with 
the long-sought evidence of a compact binary transiting from an accretion-powered 
state to a rotation-powered state (Archibald et al. 2009). Theoretical models 
(Shvartsman 1970; Burderi et al. 2001) have also suggested that these systems 
can swing between the rotation-powered and accretion-power states according to the 
mass transfer rate. Such behavior has also been firstly seen in PSR\,J1023+0038 
with the disappearance of radio pulsations (Stappers et al. 2013; Patruno et 
al. 2014), a newly formed accretion disk (Halpern et al. 2013) and the
dramatic increase of UV, X-ray and $\gamma-$ray emission (Li et al. 2014). 

In this investigation, we focus on another redback MSP \msp. 
Its spin period and spin-down rate are $P\sim1.86$~ms and 
 $\dot{P}\sim7.5\times10^{-21}$~s s$^{-1}$ respectively (Crawford et al. 2013). 
This implies a spin-down power of $\dot{E}\sim4.6\times10^{34}$~erg s$^{-1}$. 
Its distance inferred from the dispersion measure is $d\sim750$~pc, which makes
it the closest redback that has been found in the Galactic field so far 
\footnote{For updated information, please refer to 
https://apatruno.wordpress.com/about/millisecond-pulsar-catalogue/ }.

\msp\ is in a binary orbit with a period of $P_{b}\sim14.8$~hr. Its companion 
is a G-type star companion with the mass in a range of $0.4-0.7$~$M_{\odot}$ 
(Crawford et al. 2013). The orbit is almost circular with the projected 
semi-major axis of $a\sim1.23$~lt-s. The interval of radio eclipse has covered 
$\sim15\%$ of the orbit (Crawford et al. 2013). Through a long-term optical 
photometry of PSR\,J1723--2837, van Staden \& Antoniadis (2017) have found that 
the companion and the pulsar are not tidally locked.  

No $\gamma-$ray counterpart of PSR\,J1723--2837 was found in the second
\fermi\ point source catalog (2FGL; Nolan et al. 2012). Hui et al. (2014) have searched 
for the $\gamma-$ray emission from PSR\,J1723--2837 with $\sim3$~years of Fermi data and found 
a possible counterpart of $\sim6\sigma$ at the pulsar position. While this source is not in the 2FGL catalog, an unidentified source 1FGL\,J1725.5--2832 in the first \fermi\ catalog (Abdo et al. 2010) is found in the proximity of the pulsar (see Fig. 3 in Hui et al. 2014). 
Since it is $\lesssim1^{\circ}$ away from the pulsar and its $\gamma-$ray properties are 
consistent with those found by Hui et al. (2014), the authors suggest 1FGL\,J1725.5--2832 
is the same $\gamma-$ray source associated with PSR\,J1723--2837. In the latest third \fermi\
catalog (Acero et al. 2015), a source, 3FGL\,J1725.1-2832, is also found to be associated with 1FGL\,J1725.5--2832 and its flux and spectrum is fully consistent with the possible gamma-ray counterpart of \msp. 

Hui et al. (2014) have also examined the X-ray properties of PSR\,J1723--2837 with \xmm\
and \chandra. They have discovered the orbital modulation in $0.3-10$~keV with the minimum 
coincides with the phase interval of radio eclipses. Its phase-averaged X-ray spectrum is 
purely non-thermal and can be modeled by a simple power-law with $\Gamma\sim1.2$. No significant 
spectral variation across the orbit is found in this investigation. The authors have discussed 
the X-ray properties in the contest of an intrabinary shock model. Assuming a synchrotron 
origin of the X-rays, Hui et al. (2014) speculated if typical synchrotron energy is larger than 
$\sim10$~keV. However, this scenario could not be examined in their work because both \xmm\
and \chandra\ cannot provide spectral imaging data in the hard X-ray band. 

In this paper, we report a broadband X-ray analysis of PSR\,J1723--2837 with photon energies 
up to $\sim79$~keV by using the data obtained from \nustar. This is also the second redback MSP observed with \nustar\ after PSR\,J1023+0038 (Tendulkar et al. 2014; Li et al. 2014).

\section{Observations and Data Reduction}

\subsection{NuSTAR}

\msp\ was observed with \nustar\ (Harrison et al. 2013) between 2015 October 15 and 17 with an effective exposure time of $\sim 81$ ks (ObsID 30101043002). We reprocessed the raw data with the \nustar\ Data Analysis Software (NuSTARDAS) v1.7.0 under HEAsoft version 6.20 and an updated NuSTAR calibration data (CALDB version 20170120). The calibrated and and cleaned event files were produced with {\tt nupipeline}. All the data products including images, light curves, X-ray spectra and the corresponding response matrices were generated by the tool {\tt nuproducts}. 

We cleaned and filtered the events with standard parameters suggested in the \nustar\ Data Analysis Software Guide\footnote{http://heasarc.gsfc.nasa.gov/docs/nustar/analysis/nustar\_swguide.pdf}. Both focal plane modules, FPMA and FPMB, were processed separately to extract the data products. In addition to standard processing, we also performed barycentric correction with the tool {\tt barycorr} to the arrival times of all events based on the pulsar timing position at R.A. (J2000) = 17:23:23.1856, Decl. = -28:37:57.17 (Crawford et al. 2013).

The X-ray counterpart of \msp\ is clearly detected in both modules. Apart from \msp, a faint known X-ray source, 3XMM\,J172325.7--283631 (in 3XMM-DR6 catalog), is also detected at $\sim 1.5$ arcmin northeast to \msp. This faint source is nearly a factor of 10 fainter than \msp.

To extract events for spectral and temporal analysis, we used data only in the range of 3--79 keV. We employed a circular region with a radius of 50 arcsec centered at the position of \msp. A circular source-free region on the same detector chip was used for background subtraction. The X-ray spectra of the source extracted from FPMA and FPMB were rebinned to have at least 30 counts per bin.

\subsection{Swift}

Three \swift\ snapshot observations with a total exposure time of about 5.3 ks were taken simultaneously with the \nustar\ observation. In this study, we use the imaging data taken with the X-ray Telescope (XRT) onboard \swift\ to constrain the soft ($< 3$ keV) X-ray emission. We extracted the co-added X-ray spectrum (to improve the signal-to-noise ratio) and the corresponding response files by using the XRT products generator \footnote{http://www.swift.ac.uk/user\_objects} (Evans et al. 2007,2009).

\section{Data Analysis}

\subsection{Timing Analysis}

\msp\ is known to have a 14.8-hr X-ray orbital period using \chandra\ and \xmm\ observations (Hui et al. 2014). To search for the hard X-ray orbital modulation with \nustar, we combined the background subtracted light curves from FPMA and FPMB (Figure 1). In addition to the 3--79 keV light curve, we also extracted light curves from 3--10 keV, 10--20 keV, and 20--79 keV. Using the radio timing ephemeris (Crawford et al. 2013), we folded the barycentered events at the orbital period (Figure 2). The phase zero is defined as the time of the ascending node (MJD 55425.320466), meaning that the inferior conjunction (when the companion star is between the pulsar and observer) is at phase 0.25. The orbital period is clearly shown in all energy ranges concerned. The profiles of the light curves are similar to that of \chandra\ and \xmm\ (Hui et al. 2014), and the minimum of the X-ray light curves (at phase $\sim 0.25$) corresponds to the radio eclipse (Crawford et al. 2013). We also computed the ratios between different energy bands and there is no evidence of color variation throughout the orbital period (see Figure 2 for the flux ratio between 20--79 keV and 3--10 keV) and we will show in section 3.2 that the phase-resolved spectral fits of the minimum and maximum are statistically the same.

\begin{figure}
        \centering
        \psfig{file=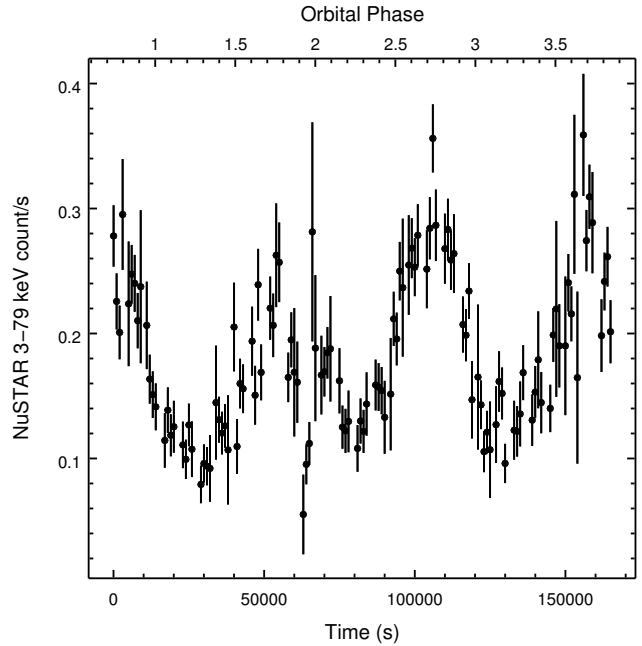,width=3.5in}
        \caption{Background subtracted \nustar\ 3--79 keV light curve of \msp. Each bin corresponds to 1000 s of exposure time.
        }
\end{figure}

Furthermore, the X-ray light curves show a dipping feature on a timescale of about 3 hours in one occasion (see Figure 1). This feature happened in between the maximum and minimum of the light curve at the orbital phase of $\sim 0.9$. To test the significance, we fit the light curve with a 14.8-hr sinusoidal plus a Gaussian centered at the dip. The addition of the Gaussian is significant at $> 99.9\%$ level by using an F-test.
We also checked the light curves in different energy bands and this feature may be seen in all bands, but it is statistically significant ($> 99.9\%$) only in the 3--10 keV band. For the other bands, the small numbers of photons do not allow us to claim a detection. The hardness ratio did not reveal any variability.

\begin{figure}
        \centering
        \psfig{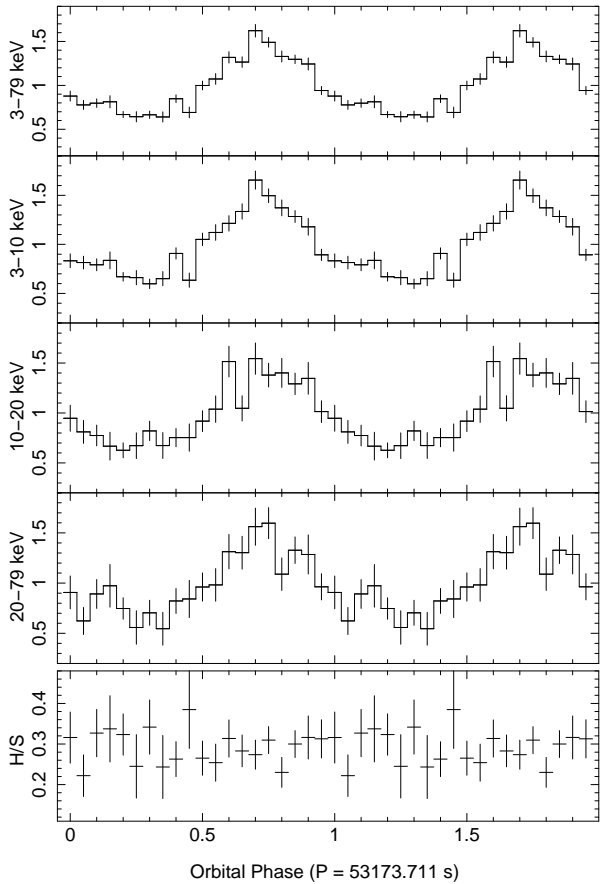}
        \caption{Folded light curves at different energy bands of \msp\ with an orbital period of 53173.711 s. From top to bottom: 3--79 keV, 3--10 keV, 10--20 keV, 20--79 keV, and hardness ratio (20-79 keV / 3--10 keV). It is clear that all light curves show the 14.8-hr orbital period. The hardness ratio remains steady throughout the orbital period, indicating that there is no color variation.
        }
\end{figure}

\subsection{Spectral Analysis}

We performed spectral analysis using XSPEC version 12.9.1. We first fit the phase-averaged spectrum with a simple absorbed power-law model. This is based on the fact that there is no indication of any thermal emission from the 0.3--10 keV spectra (Hui et al. 2014). To constrain the soft X-rays below 3 keV (where \nustar\ is not sensitive), we performed a joint fit together with the simultaneous \swift/XRT data. We also included constants to take cross-calibration between \nustar\ FPMA and FPMB and \swift/XRT into account. Figure 3 shows the broadband energy spectrum of \msp. The spectrum clearly extends up to about 70 keV with no significant emission and absorption features and can be well described by an absorbed power-law model ($\chi^2_\nu=0.87$ for 289 degrees of freedom (dof)). The best-fit model parameters are $N_H=(3.9^{+2.7}_{-2.0})\times10^{21}$ cm$^{-2}$ and $\Gamma=1.28\pm0.04$. All quoted uncertainties in this paper are 90\% confidence level. The best-fit $N_H$ is entirely consistent with the Galactic value ($\sim 4\times10^{21}$ cm$^{-2}$; Kalberla et al. 2005). The unabsorbed 3--79 keV flux is $(9.6\pm0.5)\times10^{-12}$ ergs cm$^{-2}$ s$^{-1}$, corresponding to an X-ray luminosity of $(6.5\pm0.3)\times10^{32}$ ergs s$^{-1}$ ($d=0.75$ kpc).
   
\begin{figure}
        \centering
        \psfig{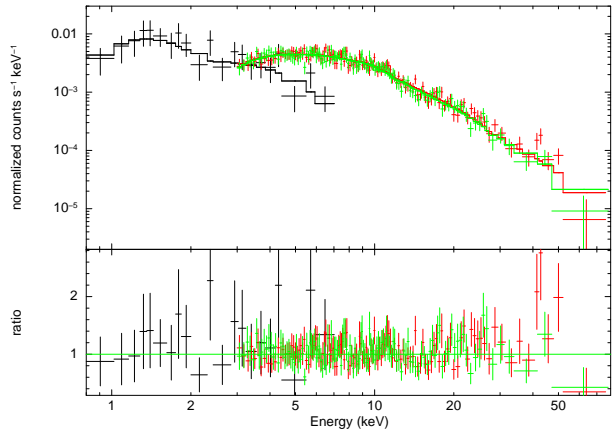}
        \caption{Broadband phase-averaged spectrum of \msp\ taken with \swift/XRT (black data points and line) and \nustar\ (red and green data points and lines). The best-fit spectrum is an absorbed power-law with $N_H=3.9\times10^{21}$ cm$^{-2}$ and $\Gamma=1.28$.
        }
\end{figure}
   
We also fit the spectrum with a power-law model with a high-energy exponential cutoff. However, the fit was not improved significantly and the cutoff energy is unreasonably large ($>300$ keV). We therefore conclude that a simple power-law model is sufficient to describe the spectrum.

We next investigated two phase-resolved spectra covering the orbital phase of 0.2--0.4 and 0.7--0.9. These two phase ranges roughly corresponds to the X-ray minimum (inferior conjunction) and maximum (superior conjunction), respectively. For the phase-resolved spectroscopy, we used the \nustar\ data only since the \swift\ observations do not have sufficient photons. We fixed the $N_H$ at the best-fit value determined from the phase-averaged spectrum. The best-fit $\Gamma$ is $1.32\pm0.12$ and $1.25\pm0.06$ for inferior and superior conjunction, respectively. Varying the $N_H$ within the 90\% uncertainty gives consistent results.
This shows that both spectra are statistically the same and are consistent with the phase-averaged spectrum. We conclude that there is no significant X-ray spectral variability throughout the binary orbit of \msp. The \nustar\ results are in general consistent with \chandra\ and \xmm\ observations but for the superior conjunction, the spectrum of \nustar\ is steeper than that of \chandra/\xmm.

\section{Discussion}

We have investigated hard X-ray (3--79 keV) properties of the MSP binary \msp\ with \nustar. The 14.8-hr binary orbital period of the system is clearly shown in hard X-ray (Figure 1 and 2) and the profile is similar to that seen in \chandra\ and \xmm\ (Hui et al. 2014). The broadband (0.3--79 keV) energy spectrum from the joint \swift\ and \nustar\ data shows an absorbed power-law with $\Gamma=1.28$ and an X-ray (3--79 keV) luminosity of $6.5\times10^{32}$ ergs s$^{-1}$. The \nustar\ results strongly support the non-thermal nature of the X-ray emission as already indicated in previous studies with \chandra\ and \xmm\ (Hui et al. 2014). To produce such a hard X-ray spectrum, it is suggested that the X-rays are from an intrabinary shock due to interaction between pulsar wind and outflow from the companion star (Takata et al. 2014; Li et al. 2014). This model also explains the X-ray orbital modulation naturally. 

The \nustar\ result is not only consistent with previous results based on soft ($< 10$ keV) X-ray data, but also similar to the transitional redback MSPs PSR\,J1023+0038. PSR\,J1023+0038 is the first redback MSP observed with \nustar\ (Tendulkar et al. 2014; Li et al. 2014). During its rotation-powered state, PSR\,J1023+0038 shows a hard spectrum ($\Gamma=1.2$) comparing to a relatively soft ($\Gamma=1.7$) spectrum seen in a low-mass X-ray binary state. The hard X-ray luminosity of the two objects during their rotation-powered state is also the same. Interestingly enough, the GeV spectra of the two sources taken with \fermi\ are also very similar with a power-law photon index of $\sim 2.5$ (Hui et al. 2014; Takata et al. 2014; Patruno et al. 2014).
With the new information provided by \nustar, we here explain the X-ray emission of \msp\ with a revised intrabinary shock model.

The orbit modulating X-rays of \msp\ are likely produced by the
 intrabinary shock like other redback MSP binary such as PSR\,J1023+0038 (Takata et al. 2014; Li et al. 2014) and PSR\,J2339--0533 (Kong et al. 2012). The observed radio eclipse at several GHz lasts only 15\% of its binary orbit, suggesting the shock wraps the companion star and its distance from the pulsar is of order of the orbit separation,
which is of order of $a\sim 5\times 10^{10}$cm (Crawford et al. 2013).
The magnetic fields at the shock may be estimated as $B_s\sim (L_{sd} \sigma/a^2c)^{1/2}\sim 7.8$G, where we used $L_{sd}=4.6\times 10^{34}{\rm erg~s^{-1}}$,
$a=5\times 10^{10}$cm and  $\sigma=0.1$ as the  the ratio of the magnetic
energy to the kinetic energy of the relativistic pulsar wind (Takata et al. 2017).
At the shock, the kinetic energy of the  pulsar wind is converted into
the internal energy, and the pulsar wind particles are accelerated beyond
the Lorentz factor of the cold relativistic pulsar wind in the upstream region.
We may estimate the maximum Lorentz factor of the accelerated particle
by balancing the accelerating time scale, $\tau_{acc}\sim \Gamma_{max}m_ec/(\zeta e B_s)$, where $\zeta<1$ is the efficiency of shock acceleration, 
and synchrotron cooling time scale, $\tau_s\sim 9m_ec^3c^5/(4e^4B_s^2\Gamma_{max})$, which yields
$\Gamma_{max}\sim 4\times 10^7\zeta^{1/2}(B_s/10{\rm G})^{-1/2}$.
These TeV electrons accelerated at the shock
will produce the very high-energy
photons by scattering off the soft photons from the companion
star ($T_{eff}\sim 4800-6000$K),
which could possibly be measured by the
Cherenkov Telescope Array (CTA) in the future. 

For the synchrotron radiation, the process is occurred in the fast-cooling regime for the shocked particles with a Lorentz factor above $\Gamma_c\sim
1.5\times 10^6 (a/5\cdot 10^{10}{\rm cm})^{-1}(B_s/10{\rm G})^{-2}$,
corresponding to the typical synchrotron photon energy of
$E_{s,c}\sim 400{\rm keV}(\Gamma_c/1.5\cdot 10^6)^2(B_s/10{\rm G})$. The
photon index changes from $\alpha=(p+2)/2$ above $E_{s,c}$ to $\alpha=(p+1)/2$
below $E_{s,c}$. At the typical synchrotron energy ($E_1$) of the minimum Lorentz factor ($\Gamma_1$) of the shocked particles, the photon index changes
from $\alpha=(p+1)/2$ above $E_1$ to $\alpha=2/3$ below $E_1$. \chandra,
\xmm, and \nustar\ observations revealed that the X-ray emissions in
0.3--79 keV  energy range is well described by a simple power-law with a
photon index of $\sim 1.2-1.3$, which is
similar to the transitional MSP PSR\,J1023+0038.
This simple power law spectrum may indicate that the typical synchrotron
energy $(E_1)$ of the minimum Lorentz factor $\Gamma_1$ is smaller than
0.1 keV, and the power law index of the particle distribution at the shock is
$p\sim 1.4-1.6$. In such a case, the minimum Lorentz factor is constrained as
$\Gamma_1\le 2.4\times 10^4(B_s/10{\rm G})^{-1/2}$.

Apart from the X-ray orbital modulation, the X-ray light curve also shows a short-term dipping feature on a timescale of about 3 hours in between the inferior and superior conjunctions. Given the orbital phase ($\sim 0.9$) of the dip, it is unlikely due to an eclipse. The lack of a hardness change during the dip indicates that it may be intrinsic to the X-ray radiation (i.e., intrabinary shocked X-rays) caused by perturbation of shock front via clumpy pulsar wind or outflow variation from the companion. For instance, if the outflow velocity from the companion has a sudden drop, the momentum ratio of the stellar wind and the pulsar wind will drop accordingly, resulting a smaller fraction of the pulsar wind stopped by the outflow.

In summary, we found modulating X-ray radiation at the 14.8-hr binary orbit of \msp\ up to 79 keV with \nustar. The energy spectrum is a simple power-law with a photon index of $\sim 1.3$ and shows no evidence of spectral variability throughout the orbital cycle. Such radiation can be explained with an intrabinary shock model.

AKHK is supported by the Ministry of Science and Technology of the Republic of China (Taiwan) through grants 105-2112-M-007-033-MY2, 105-2119-M-007-028-MY3, and 106-2918-I-007-005. CYH is supported by the National Research Foundation of Korea through grants 2014R1A1A2058590 and 2016R1A5A1013277. JT is supported by NSFC grants of Chinese Government under 11573010 and U1631103.
   
{\it Facilities:} \facility{NuSTAR}, \facility{Swift}

\end{document}